# Detailed ROSAT X-ray Analysis of the AM Her Cataclysmic Variable VV Pup

E. El kholy[1,2] and M. I. Nouh[1,2].

[1]Physics Department, College of Science, Northern Border University

1320, Arar, Saudi Arabia

[2]Departmt of Astronomy, National Research Institute of Astronomy and Geophysics

11421, Helwan, Cairo, Egypt

**ABSTRACT:** VV Pup is typical system of AM Her stars, where the main accreting pole rotates in and out of view during the orbital cycle. In the present paper we present ROSAT data analysis for the magnetic cataclysmic variable VV Pup. We obtained the X-ray light curves of VV Pup in high state, the PSPC count rate 0.1-2.0 keV is plotted as a function of time with bins of 10 sec and the count rate is folded over the orbital period of 100.4 min with bin size of 100 sec for individual observations. We calculate the mean best-fit PSPC spectrum, with a three-component spectral fit including a soft X-ray blackbody, hard X-ray bremsstrahlung, and Gaussian line covers the phase intervals, for a bright phase($\varphi_{orb}$=0.9-1.1), the dip data($\varphi_{orb}$=0.18-0.7), the egress from the dip ($\varphi_{orb}$=0.7-0.8), the phase interval $\varphi$ = 0.1-0.18 and the mean best-fit spectrum for all data ($\varphi_{orb}$=0.0-1.0). We calculate spectral parameters, the hardness ratios, count rate and total integrated black body flux.

1. Introduction

The Magnetic Cataclysmic variables (MCVs) are ideal plasma laboratories in which magneto-hydro dynamical problems and radiation processes for matter under extreme conditions (temperatures of about 100 M k and Magnetic field about 100 MG, Burwitz (1997).
The Cataclysmic variables (CVs) are mass-transfer binary systems that consist of a normal star (the secondary) and a white dwarf (the primary) with an orbital period in the range 1-10 hrs. A white dwarf accretes material from a late type main-sequence star through Rouch lobe overflow. If the white dwarf has a significant magnetic field, the white dwarf rotates synchronous with the orbital period (Magnetic Cataclysmic variables or Polars). VV Pup is one of the AM Herculis system, it has two poles 31.5 MG and 54.6 MG (Mason, 2007).

AM Herculis (Polars), is the prototype of the class of synchronized magnetic CVs, The magnetic field strength at the usually dominant accreting, The strong magnetic field of the white dwarf completely controls the accretion by preventing the formation of accretion disc and channeling the accreting matter along the field lines. The accreting matter reaches supersonic velocities and encounters shocks near the white dwarf surface. The shocks heat up the infilling matter to high temperatures. The post shock hot plasma cools as it plunges towards the white dwarf emitting hard x-rays via thermal bremsstrahlung, Girish et al. (2006).

VV Pup is the proto type of the class of synchronized magnetic CVs, it is identified as the third example of the AM Her (Topia, 1977) with orbital period of 100.4 min (Walker, 1965), the field strength of VV Pup in first pole $B_1=30.5$ MG and in second pole $B_2=54$ MG (Schwope, 1997). Ramsay et al. (1996) found significant spectral variation in the soft x-ray component (black body model).

In the present paper we are going to perform light curve and spectral analyses of VV Pup. Count rates are obtained at different phases. We fitted the spectra with different models to explore the spectral variation of the system. The structure of the paper is as follows: Section 2 is devoted to the observations and data analysis. Section 3 deals with the x-ray light curves. In section 4 we perform the spectral analysis of the object. Section 5 outline the conclusion reached.

2. **Observations and Data Analysis**

The VV Pup is one of AM Herculis system (Polars). The VV Pup was observed with the ROSAT satellite X-ray data from the Position Sensitive Proportional Counters (PSPC) operating the 0.1-2.4 keV spectral band on 17 October 1991 with the total exposure times 17037 sec, the average count rate for all channels (energies) is 10.33 cts/sec. The PSPC spectrum was analyzed using the detector response matrix DRMPSPC_B01C, Beuermann et al. (2008). Table 1 lists ROSAT observations of VV Pup.

Table 1: ROSAT Observations of VV Pup.

| Date of observation UT | INSTRUMENT_NAME | Average Count rate for all energies (cts/s) | Exposure (Sec) | Observation Id | RA/DEC (2000) |
|---|---|---|---|---|---|
| 17-OCT-1991 07:57:29 | PSPC B | 10.33 | 17038.29 | WG300140P | 08h 15m 06. 8s -19d 03m 18.0s |

Figure 1 shows the PSPC images of the Oct. 1991 high state observations of VV Pup, the images size are 42 x 42 arc min and the pixel size is 5 arc sec, with background sources removed , the inner circle for the source ($r_s$ = 4xFWHM, 4 arc min) , the background is obtained by collecting in an outer annulus, it is between the second circle (the inner radii $r_i$ = 5 FWHM , 5arc min ) and the outer circle (r = 12 FWHM , 12 arc min).  In Figure 1, panel a, the image for all pulse channels of photons between 0.05 to 2.4 keV with bin image 10 sec.  Panel b, the image dominated by soft X-ray with photons energies 0.05 and 0.12 Kev from channels 5 to 12. Panel c, the image shows that the x-ray photons energies from channels 20 to 50 cover energies 0.20 to 0.50 keV. Panel d, represents the hard x-ray photons energies which cover channel 50 to channel 200 (energies from 0.5 to 2.0 keV). One can see a light spot under the center of the hard source, it seems as another source but after careful selection of the photons energies, it disappears.

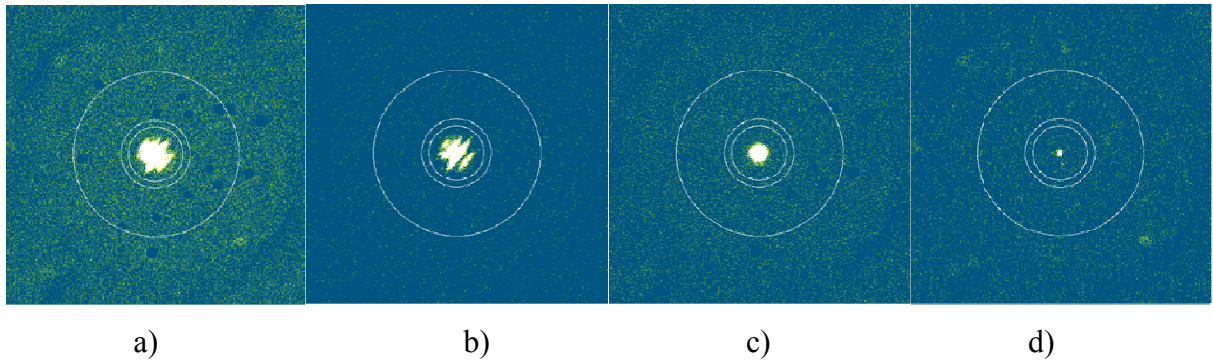

a)                               b)                               c)                               d)

**Figure 1**: PSCP images of the October 1991 high state observation of VV Pup with background sources removed, The image size is 42x42 arc min, panel a: The image for all photons energies between 0.01 to 2.4 Kev, panel b:  the image shows the soft selected X-ray photons energies between 0.05 and 0.12 Kev, panel c: the image for X-ray photon energies range 0.2-0.5 keV, and panel d: the image for hard X-ray photon energies range 0.5-2.0 keV.

The X-ray light curve of VV Pup in its high state is shown in Figure 2, where the PSPC count rate in channels 10 -200 (0.10-2.0 keV) is plotted as function of time bins of 10 sec. It represents the PSPC data taken on 17 October 1991 with the total exposure time of 17037 sec. broken up into 16 interval observations (hereafter we will call it OBIs). The mean count rate is 10.33 cts/s. OBIs #4,#7,#15 and part of OBIs #8 cover the self-eclipse when the accretion spot disappears behind the limb of the white dwarf. Also Figure 2 shows that substantial variability exists between the individual OBIs during the high state observation. The count rate in OBIs #4, #7 and #15 with exposure times of 1291, 1089 and 814 sec are very low, and the count rate close to the count rate of the background, the OBIs #8, #10, #11 with exposure times of 1564, 1643 and 636 sec are comparatively low (~ 12 cts s$^{-1}$) and the counts rate are high in OBIs #2, #3 and #1 reaches a maximum of 100 cts s$^{-1}$ per 10 sec bin.

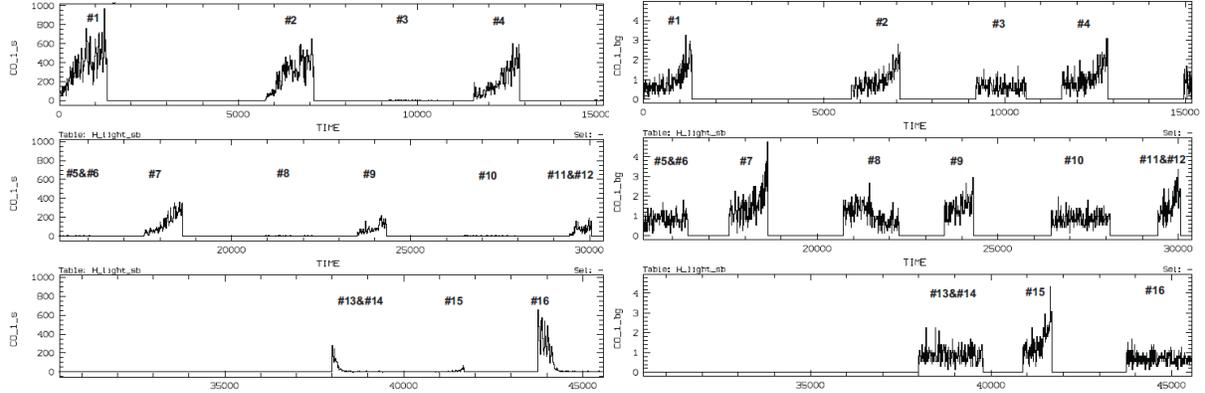

**Figure 2:** X-ray light curve PSPC data of VV Pup in high state, taken on 17 October 1991, per 10 sec bin, with the total exposure time of 17037 sec. Left panel: the count rate of source vs. time as observed and broken up into 16 observation intervals (OBIs). Right panel represents the background.

## 3. X-ray Light Curves

The X-ray orbital light curve of VV Pup on October 1991 high state is shown in Figure 3 with a total exposure time 17037 sec, the count rate is folded over the orbital period of 100.4 min and with time bins 10 sec, using the ephemeris of Patterson et al. (1984) given in equation (1).

$$\text{Maximum light} = JD\odot 2427889.6474 + 0.0697468256E \qquad (1)$$

Figure 3 shows the Orbital light curve of VV Pup, The count rate of source versus the orbital phase with individual OBIs shown in Figure 2. The bright phase (br0) covers roughly the phase interval $\varphi_{orb} = 0.9 - 1.1$ and includes the OBIs 1, 2, 4, 7 and 16 and indicated by different color symbols. The mean count rate of the bright phase (br0) is 92.59 cts/sec with total time 1911.7 sec. The dip data (br4) between the phase interval of $\varphi_{orb} = 0.18 - 0.70$, the count rate of the dip is very low 1.112 cts/s. The dip phase interval includes part of the OBIs 3, 5, 6, 8 and 10 to16 with total time 9312.43 sec. The accretion spot has the best view in the bright phase interval br0 ($\varphi_{orb}$=0.9 –1.1), where the dips behind the limb of the white dwarf. The OBIs 2, 4, 7, 9, 11, 12 and 15 which cover the phase interval ($\varphi_{orb}$ = 0.7–0.8) represents the egress from the dip (br1) with an intermediate count rate of 7.148 cts/s and the total time reach to 2297 sec. OBIs 1 and 2 contain a large flare (accretion event) which reaches a peak count rate ~100 cts/s at resolution of 10 sec. The phase intervals br2 ($\varphi_{orb}$ = 0.8–0.9) and br3 ($\varphi_{orb}$ = 0.1–0.18) are before and after the bright phase, the count rate are 21.07 cts/s and 13.84 cts/s respectively. We show a summary of the results in Table 2.

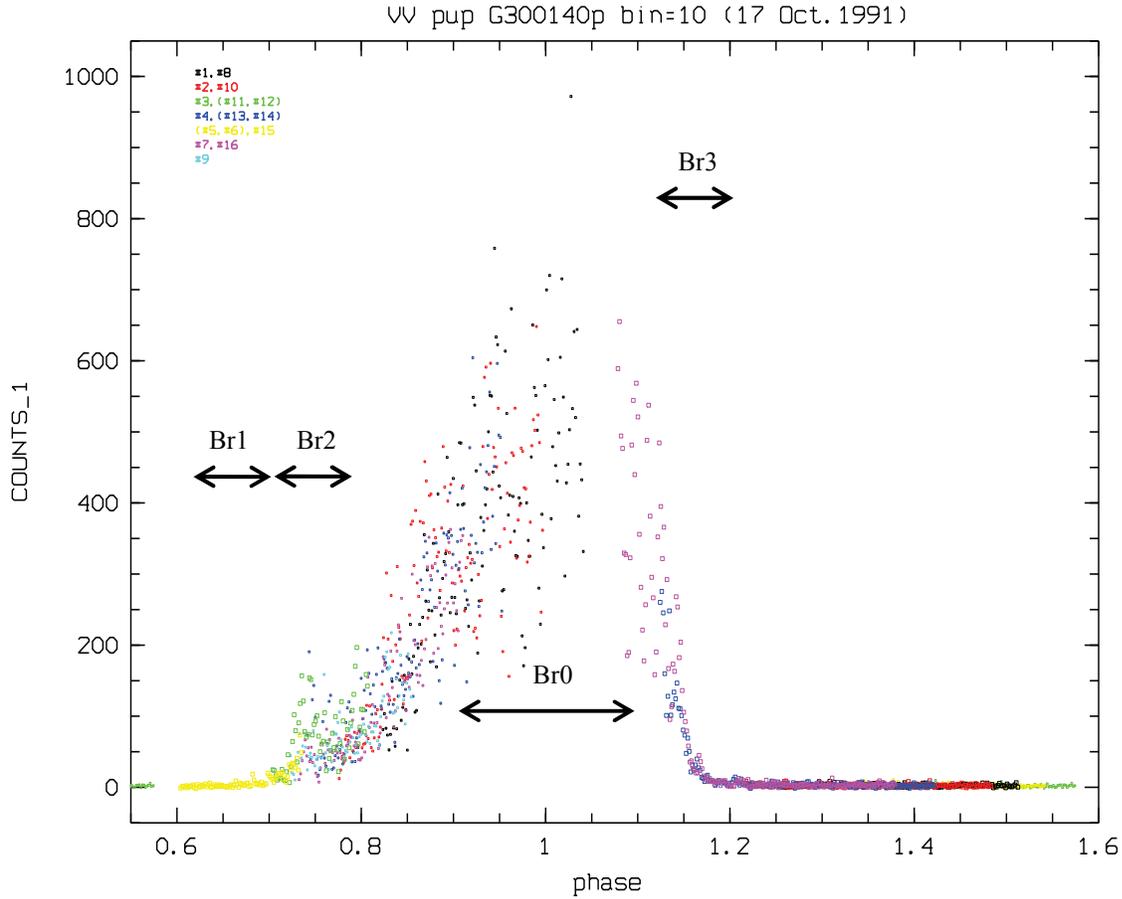

**Figure 3 :** The PSPC count rate in channels 1-240 (0.01-2.4keV) folded on the orbital period of 100.4 min with a bin size 10 sec, using the ephemeris of Patterson et al. (1984). The horizontal bars labelled br0, br1, br2 , br3 and br4 indicate phase intervals $\varphi_{orb}$ = 0.9–1.1 for bright phase (br0), $\varphi_{orb}$ = 0.7–0.8 for egress from the dip (br1), $\varphi_{orb}$ = 0.8–0.9 for interval before the bright phase (br2), $\varphi_{orb}$ = 0.1–0.18 for interval after the bright phase (br3) and $\varphi_{orb}$ = 0.18–0.7 for the dip interval (br4).

**Table 2:** Results for the phase intervals T0($\varphi$ = 0.0–1.0 for all OBIs), Br0 for bright phase, br1 for the egress from the dip, br2, br3 and br4 for the dip ($\varphi$ = 0.18–0.7) with exposure times, number of photons and count rate.

| Phase interval | $\Phi_{orb}$ | OBIs | No. of photons | Total time Sec. | Count rate |
|---|---|---|---|---|---|
| T0 | 0.0-1.0 | 1 to 16 | 176083 | 17038.29 | 10.33 |
| Br0 | 0.9-1.1 | 1,2,4,7,16 | 81424 | 1911.7 | 92.59 |
| Br1 | 0.7-0.8 | 2,4,7,9,11,12,15 | 16426 | 2297.83 | 7.15 |
| Br2 | 0.8-0.9 | 1,2,4,7,9,12 | 56017 | 2659.24 | 21.07 |
| Br3 | 0.1-0.18 | 13,16 | 11858 | 857.08 | 13.84 |
| Br4 | 0.18-0.7 | 3,5,6,8,10-16 | 10358 | 9312.43 | 1.11 |

We calculate the hardness ratios HR1 (Equation 2), HR2 (Equation 3) and the count rate ratio S1/S2 (Equation 4) of the two energy intervals in the soft band by the following equations.

$$HR1 = (HARD - SOFT) / (HARD + SOFT) \qquad (2)$$
$$HR2 = (SOFT2 - SOFT1) / (SOFT2 + SOFT1) \qquad (3)$$
$$S1/S2 = SOFT1/SOFT2 \qquad (4)$$

Where SOFT and HARD are the count rates in the standard intervals of 0.11-0.41keV and 0.52-2.01keV respectively, we use the ratio HR2 for which the two bands SOFT1 and SOFT2 are 0.10-0.20 keV and 0.20-0.40 keV, respectively.

We show in Figure 4 the hardness ratios HR1, HR2, the count rate ratio S1/S2 of two energy intervals in the soft band and the count rate as a function of phase ($\varphi$) with time bins 100 sec. We found some correlations between these hardness ratios versus count rate. The mean value of HR1 increases in the dip ( br4 interval, $\varphi_{orb}$=0.18-0.7), this means that the source becomes harder when it is close to the horizon of the white dwarf. The hardness ratio HR1 increases near $\varphi$ = 0.18 where the count rate drops and also increases in begin the egress from the dip near br1 $\varphi$ = 0.7.

The flux tends to become softer at the bright phase interval $\varphi_{orb}$=0.9-1.1. Also we found some correlation is immediately evident, the source become harder at $\varphi$=0.7, then it will tends to be softer at $\varphi$=0.75 with increase the count rate , then the source back again harder with increasing count rate.

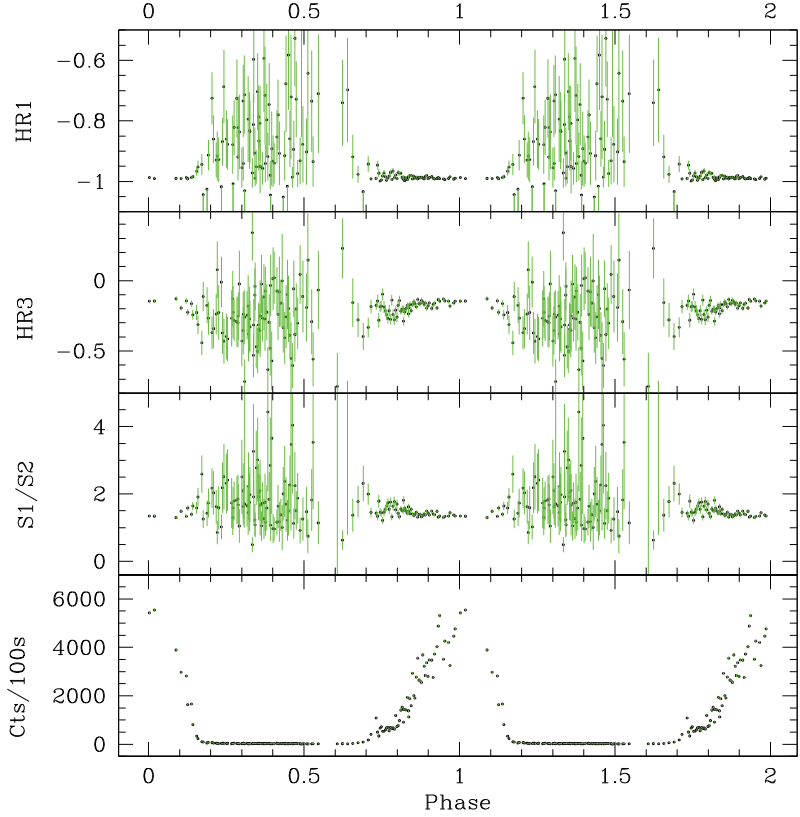

**Figure 4:** Orbital light curve of VV Pup in the October 1991 high state as in Figure 2. The individual panels from top to bottom give the hardness ratios HR1 and HR2, the count rate ratio S1/S2, and the total count rate folded on the orbital period of 100.4 min with a bin size 100 sec.

## 4. X-ray Spectra

The Position Sensitive Proportional Counter PSPC (Pfeffermann et al., 1986) sensitivity is in the range of 0.1-2.4 Kev. We used MIDAS software and EXSAS software package to prepare the spectrum for timing analysis, spectral analysis and spectral modeling.

Spectral analysis includes the three selected phases intervals, shown in Figure 3, phase interval T0 ($\varphi_{t0}$=0.0-1.0), bright phase interval Br0 ($\varphi_{Br0}$=0.9-1.1), and phase interval Br1 (egress from the dip $\varphi_{Br1}$=0.7-0.8) respectively. We fit the total phase, (T0, $\varphi_{t0}$=0.0-1.0), spectrum of VV Pup with a three component model, namely soft X-ray blackbody component plus hard X-ray thermal bremsstrahlung component, and an intermediate component Raymond-Smith thermal or a simple Gaussian line.

In order to investigate the dependence of the spectral parameters on orbital phase, spectra were accumulated over the total phase T0 ($\varphi$=0.0-1.0), the egress from the dip br1 ($\varphi$=0.7-0.8) and the bright phase Br0 ($\varphi$=0.9-1.1).

Figure 5a, describes the mean spectrum of a 17038 sec of all OBIs with count rate 10.33cts s$^{-1}$, the total spectrum is fitted well with a three component model of blackbody, thermal bremsstrahlung and Gaussian line. The best fitted parameters are, kT$_{brems}$=20 keV fixed, the best column density N$_H$=1.0 x 10$^{18}$ cm$^{-2}$, the best blackbody temperature kT$_{bb}$= 32.2 keV, the Gaussian line energy at 0.85 keV and its FWHM fixed at 0.1keV and $\chi^2$=0.97. The integrated X-ray blackbody fluxe is F$_{bb}$=5.44x10$^{-11}$ erg cm$^{-2}$ s$^{-1}$.

Figure 5b shows the spectrum of a 16424 sec, the egress from the dip br1($\varphi$=0.7-0.8), taking in part from OBIs 2,,4,7,9,11,12 and 15, with count rate 7.15cts/sec. The br1 phase spectrum is fitted with three component model of blackbody, thermal bremsstrahlung and Gaussian line, with kT$_{brems}$=20 keV fixed. The best fit of kT$_{bb}$=28.5 keV, the best column density N$_H$=1.0 x 10$^{18}$ cm$^{-2}$, $\chi^2$=0.61 and the total integrated X-ray blackbody flux is F$_{bb}$=4.31x10$^{-11}$ erg cm$^{-2}$ s$^{-1}$.

Figure 5c depicts the br1 phase spectrum fitted with a three component model of blackbody, thermal bremsstrahlung and Raymond Smith thermal, with kT$_{brems}$=20 keV fixed, the column density N$_H$=1.0 x 10$^{19}$ cm$^{-2}$ fixed, the best blackbody temperature kT$_{bb}$= 27.0 keV, the best fit thermal Raymond Smith temperature kT$_{rams}$=0.72 keV with $\chi^2$=0.62. The integrated X-ray blackbody fluxes F$_{bb}$=5.64x10$^{-11}$ erg cm$^{-2}$ s$^{-1}$.

Figures 5d, e, f and g show the PSPC spectra in the bright phase Br0 ($\varphi$=0.9-1.1). In Figure d, the spectrum is fitted with three component model of (bb+th+gl) with kT$_{breams}$=20 keV fixed, the best fit column density N$_H$=5.5 x 10$^{18}$ cm$^{-2}$, the best blackbody temperature kT$_{bb}$= 32.3 keV, the best fit gives $\chi^2$=0.88 and the integrated X-ray blackbody fluxes F$_{bb}$=2.51x10$^{-10}$ erg cm$^{-2}$ s$^{-1}$. In Figure 5e the spectrum is fitted with the three component of (bb+th+rs), the best blackbody temperature is kT$_{bb}$=25.6 keV , the best N$_H$= 5.5 x 10$^{18}$ cm$^{-2}$, the thermal Raymond Smith temperature kT$_{rams}$=0.30 keV fixed. Figures f and g indicate the spectra fitted with three component model of (bb+th+gl) and (bb+th_rs) with kT$_{breams}$=20 keV fixed, the column density N$_H$=1.0 x 10$^{19}$ cm$^{-2}$ fixed, the best blackbody temperature (Figure f: kT$_{bb}$= 31.3eV, F$_{bb}$=2.83x10$^{-10}$ erg cm$^{-2}$ s$^{-1}$ ,$\chi^2$=0.98) and (Figure g: kT$_{bb}$=31.4eV, F$_{bb}$=2.81x10$^{-10}$ erg cm$^{-2}$ s$^{-1}$, $\chi^2$=0.87).

**Table 3:** Summary of the X-ray spectral fit parameters of the high state of VV Pup in October 1991 to the integrated spectra using the sum of three models, an absorbed blackbody (free $kT_{bb}$), Thermal bremsstrahlung ($kT_{brems}$ =20keV Fixed), and an intermediate component Raymond-Smith thermal (rs) or a simple Gaussian line (gl).

| Phase $\Phi_{orb}$ | fit | Model | $N_H$ ($10^{21}$/cm$^2$) | Norm$_{bb}$ | $kT_{bb}$ (eV) | $kT_{th}$ (keV) fixed | $\chi^2_\nu$ | Count rate |
|---|---|---|---|---|---|---|---|---|
| T$_0$(0.0-1.0) | a | bb+gl+tb | 1.0e-3 | 3.93e-1 | 32.0 | 20 | 0.972 | 10.33 |
| Br1(0.7-0.8) | b | bb+gl+tb | 1.0e-3 | 3.50e-1 | 28.5 | 20 | 0.612 | 7.15 |
| Br1(0.7-0.8) | c | bb+rs+tb | 1.0e-2 fixed | 4.83e-1 | 27.0 | 20 | 0.628 | 7.15 |
| Br0(0.9-1.1) | d | bb+gl+tb | 5.55e-2 | 1.80e+0 | 32.3 | 20 | 0.883 | 92.59 |
| Br0(0.9-1.1) | e | bb+rs+tb | 6.37e-2 | 8.35e+0 | 25.6 | 20 | 1.190 | 92.59 |
| Br0(0.9-1.1) | f | bb+gl+tb | 1.00e-2 fixed | 2.09e+0 | 31.3 | 20 | 0.985 | 92.59 |
| Br0(0.9-1.1) | g | bb+rs+tb | 1.00e-2 fixed | 2.07e+0 | 31.4 | 20 | 0.870 | 92.59 |

In order to further exploration the temperature range and neutral hydrogen column density range $N_H$, we have performed a grid search of the $\chi^2$-palne ($N_H$-$kT_{bb}$), as confidence contours shape "banana diagram", it has a strong correlation between $kT_{bb}$ and $N_H$, the contour lines indicate the 1σ (68.3%), 2σ (95.5%), 3σ(99.7%), 4σ(99.9%) and 5σ(100%) confidence levels, Figure 6, the confidence contour lines give blackbody temperature range $kT_{bb}$=30-33.8 keV, column density rang $N_H$=0-2x10$^{19}$cm$^{-2}$ for 1σ (68.3%), kt$_{bb}$=26-34 eV, $N_H$=0-3x10$^{19}$cm$^{-2}$ for 2σ (95.5%) and kt$_{bb}$=22.6-35 eV, $N_H$=0-6.8x10$^{19}$cm$^{-2}$ for 5σ (100%).

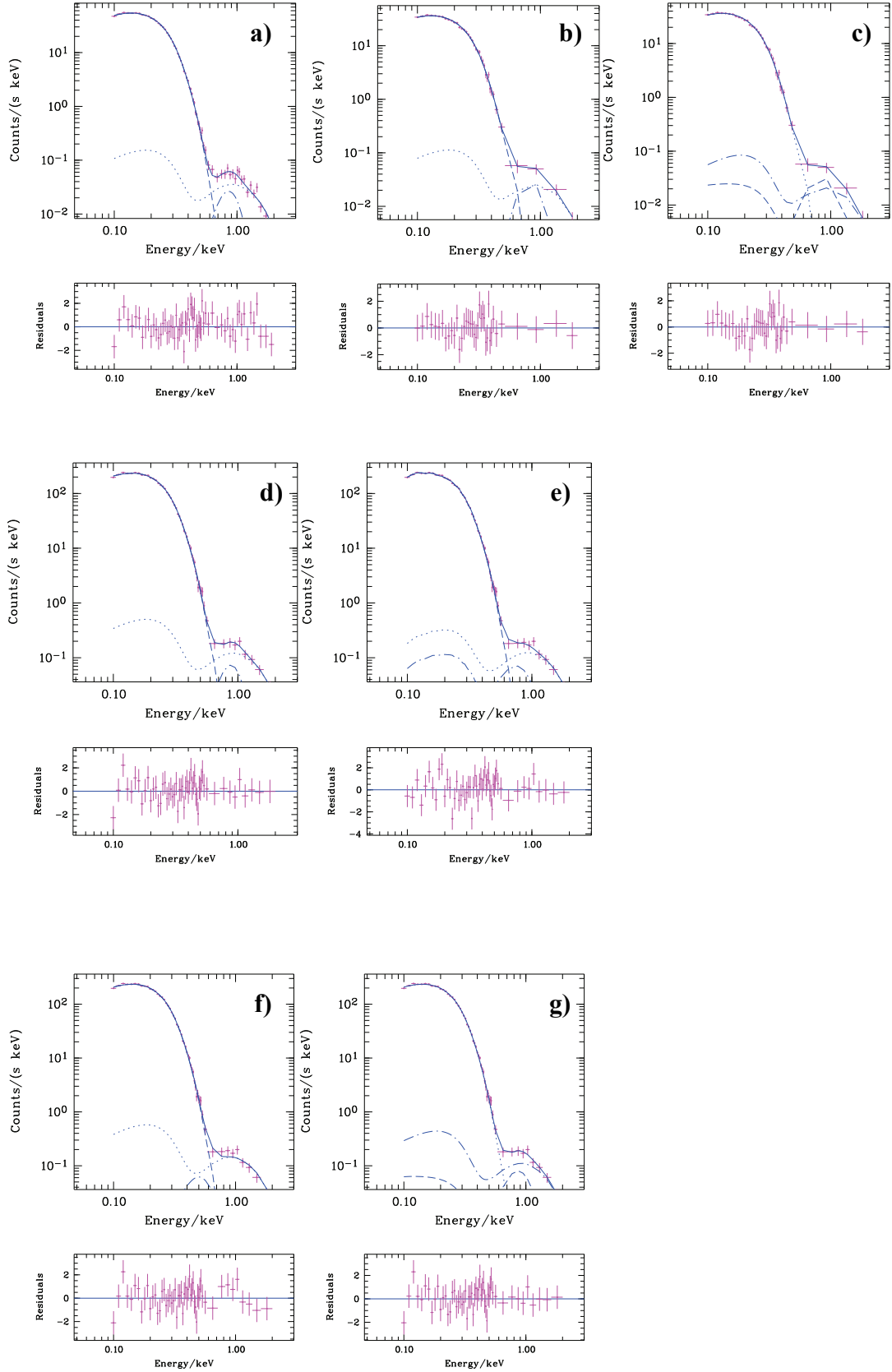

F**igure 5**: The PSPC spectra extracted from the three phase intervals, $T_0$ for all time, Br0 for bright phase and br1 the egress from the dip, the best fit PSPC spectra with a three multi-models, soft X-ray blackbody (bb), hard X-ray thermal bremsstrahlung (brems), and an intermediate component Raymond-Smith thermal (rs) or a simple Gaussian line (gl). Panel a: Best three-component spectral fit (bb+gl+th) for total phase T0. Panels b and c: Best three-component spectral fits (bb+gl+th) and (bb+rs+th) respectively through phase interval br1 (see Table 3). Panels d, e, f and g: Best three-component spectral fits (bb+gl+th), (bb+rs+th), (bb+gl+th, with fixed $N_H$) and (bb+rs+th, with fixed $N_H$) respectively for bright phase Br0.

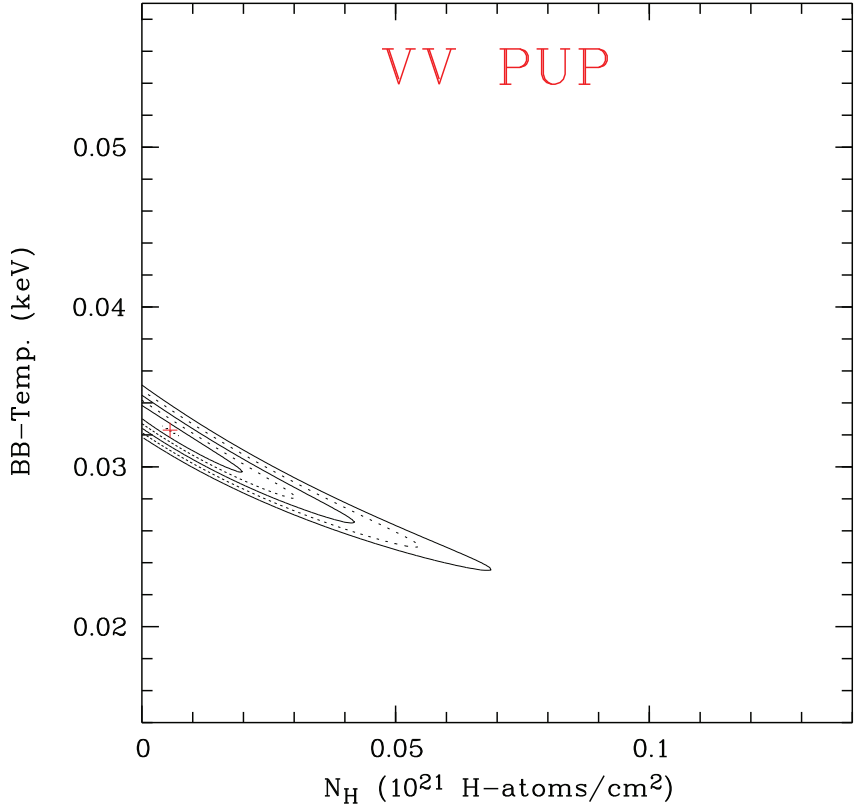

Figure 6: Result of the grid search in the $T_{bb}$-$N_H$ plane, the "banana" diagram. The 1σ, 2σ, 3σ, 4σ and 5σ contours are shown.

## 5. Discussions and Conclusions

We present ROSAT data analysis for the magnetic cataclysmic variable VV Pup. We obtained the X-ray light curves of VV Pup in high state. For individual OBIs, the PSPC count rate 0.1-2.0 keV is plotted as a function of time with bins of 10 sec and the count rate is folded over the orbital period of 100.4 min with bin size of 100 sec. We calculate the mean best-fit of the PSPC spectrum with a three components, namely a soft X-ray blackbody, hard X-ray bremsstrahlung and Gaussian line cover different phase intervals, bright phase ($\varphi_{orb}$=0.9-1.1), the dip data ($\varphi_{orb}$=0.18-0.7), the egress from the dip ($\varphi_{orb}$=0.7-0.8) and all data ($\varphi_{orb}$=0.0-1.0). We calculate the hardness ratios HR1, HR2, S1/S2, count rate and total integrated black body flux. The results reached could be drawn through the following points:

- The light curve of high state VV Pup shows that, real X-ray variability exists in OBIs 1,2,4 and 16, it is clear that the x-ray flux increases with time in OBIs 1,2 and 4 (egress from the dips, Br2 $\varphi$=0.8,0.9). The bright phase interval (br0 $\varphi$=0.9, 1.1), is the best view of the accretion spot (Figures 2 and 3), the short fluctuation flares are due to gas packages (blobs). The OBIs 3, 5, 6, 8, 10 and 15 cover the eclipse (the dip $\varphi$=0.18-0.7) when the accretion spot disappears behind the limb of the white dwarf. The shape of X-ray light curve of VV Pup is in good agreement with the differential light curve of optical observation of Hoard (2002) and light curve from the EUVE DE telescope (Stephane, 1993).

    We plotted the correlations of the hardness ratios HR1, HR2 and R=s1/s2 versus count rate as a function of phase ($\varphi$) with time bins 100 sec. The mean value of HR1 increases in the dip ($\varphi_{orb}$=0.18-0.7), this means that the source become harder when it is close to the horizon of the white dwarf. The increasing of the hardness ratio HR1 near $\varphi$=0.18 is due mainly to the increasing of the photoelectric absorption which reduces the flux of soft photons more than that of harder photons or it could be that the temperature increase when the source becomes fainter. The count rate reaches to 600 cts s$^{-1}$ in flare at $\varphi_{orb}$=1, this means that the flux still tends to become softer when the source is in the bright phase.

- The spectral analysis is done using three different spectral components, a soft X-ray blackbody, a hard X-ray bremsstrahlung component, and an intermediate component (Raymond Smith thermal or simple Gaussian line).

- The blackbody temperatures are high in bright phase spectra, this is due to high accretion rate, tends to be associated with a high density of the accreted matter which penetrates deeper into the white dwarf atmosphere and heats it from deep layer, giving rise to soft X-ray flux.